\documentclass[conference]{IEEEtran}
\IEEEoverridecommandlockouts
\usepackage{amsmath,amsfonts}
\usepackage{algorithmic}
\usepackage{algorithm}
\usepackage{array}
\usepackage{textcomp}
\usepackage{stfloats}
\usepackage{url}
\usepackage{verbatim}
\usepackage{graphicx}
\usepackage{cite}
\usepackage{amssymb}
\usepackage[utf8]{inputenc}
\usepackage{xcolor}
\usepackage[caption=false,font=normalsize,labelfont=sf,textfont=sf]{subfig}
\def\BibTeX{{\rm B\kern-.05em{\sc i\kern-.025em b}\kern-.08em
		T\kern-.1667em\lower.7ex\hbox{E}\kern-.125emX}}
\newcommand{\norm}[1]{\left\lVert#1\right\rVert}

\usepackage[top=0.75in, bottom=1.18in, left=0.635in, right=0.635in]{geometry}

\begin{document}
	
	% To decrease space above and below equations
	\setlength{\abovedisplayskip}{2.6pt}
	\setlength{\belowdisplayskip}{2pt}
	
	\title{Remote Interference Mitigation through Null Precoding and Fractional Programming}
	
	\author{Xuyang Sun\IEEEauthorrefmark{1}, Hussein A. Ammar\IEEEauthorrefmark{2}, Israfil Bahceci\IEEEauthorrefmark{3}, Raviraj Adve\IEEEauthorrefmark{1}, Gary Boudreau\IEEEauthorrefmark{3}, Zehua Li\IEEEauthorrefmark{1}
		\\
		\IEEEauthorrefmark{1}Dept.~of Elec.~and Comp.~Eng., University of Toronto, Toronto, ON, Canada\\
		
		\IEEEauthorrefmark{2}Dept.~of Elec.~and Comp.~Eng., Royal Military College of Canada, Kingston, ON, Canada\\
		\IEEEauthorrefmark{3}Ericsson Canada, Ottawa, ON, Canada
		% Email: rsadve@ece.utoronto.ca; hussein.ammar@rmc-cmr.ca; israfil.bahceci@ericsson.com;\\
		% gary.boudreau@ericsson.com; samzehuali.li@mail.utoronto.ca; eceym.liu@mail.utoronto.ca; marcoxuyang.sun@mail.utoronto.ca;
	}
	
	%\author{\IEEEauthorblockN{1\textsuperscript{st} Given Name Surname}
		%\IEEEauthorblockA{\textit{dept. name of organization (of Aff.)} \\
			%\textit{name of organization (of Aff.)}\\
			%City, Country \\
			%email address or ORCID}
		%\and
		%\IEEEauthorblockN{2\textsuperscript{nd} Given Name Surname}
		%\IEEEauthorblockA{\textit{dept. name of organization (of Aff.)} \\
			%\textit{name of organization (of Aff.)}\\
			%City, Country \\
			%email address or ORCID}
		%\and
		%\IEEEauthorblockN{3\textsuperscript{rd} Given Name Surname}
		%\IEEEauthorblockA{\textit{dept. name of organization (of Aff.)} \\
			%\textit{name of organization (of Aff.)}\\
			%City, Country \\
			%email address or ORCID}
		%\and
		%\IEEEauthorblockN{4\textsuperscript{th} Given Name Surname}
		%\IEEEauthorblockA{\textit{dept. name of organization (of Aff.)} \\
			%\textit{name of organization (of Aff.)}\\
			%City, Country \\
			%email address or ORCID}
		%\and
		%\IEEEauthorblockN{5\textsuperscript{th} Given Name Surname}
		%\IEEEauthorblockA{\textit{dept. name of organization (of Aff.)} \\
			%\textit{name of organization (of Aff.)}\\
			%City, Country \\
			%email address or ORCID}
		%\and
		%\IEEEauthorblockN{6\textsuperscript{th} Given Name Surname}
		%\IEEEauthorblockA{\textit{dept. name of organization (of Aff.)} \\
			%\textit{name of organization (of Aff.)}\\
			%City, Country \\
			%email address or ORCID}
		%}
	
	\maketitle
	
	\begin{abstract}
		With the rapid deployment of 5G systems, remote interference (RI) caused by atmospheric ducting has emerged as an occasional, but critical challenge. This phenomenon occurs when the downlink (DL) signals from distant base stations (BSs) propagate over long distances through tropospheric ducting, severely disrupting uplink (UL) reception at local BSs. To address this challenge, we analyze the effect of RI on network performance, including the channel estimation phase. We then develop a solution that identifies the angle-of-arrival (AOA) estimation of RI and designs precoders and combiners that mitigate RI. Our approach employs interference cancellation techniques through null precoding and fractional programming which enhance the performance of the network. Interestingly, we show that using our scheme, uplink communication is possible at low transmit power regimes that were unusable due to RI. Our results further show a 5.23~dB reduction in normalized mean square error for channel estimation and achieved data rates around 5.8~bit/s/Hz at the previously unusable low uplink transmit power conditions.
	\end{abstract}
	
	
	\section{Introduction}
	Remote interference (RI) is an undesired phenomenon that exists in today's wireless cellular networks \cite{RI observed}. It remains an underexplored topic by the research community despite its significant impact on the performance of wireless networks. For example, according to~\cite{power comparison}, China Mobile's field data showed 27.6\% of Xuzhou cells experienced significant uplink interference during RI, causing potential signal outages. 
	
\begin{figure}[t]
		\centering
		\includegraphics[width=0.9\linewidth]{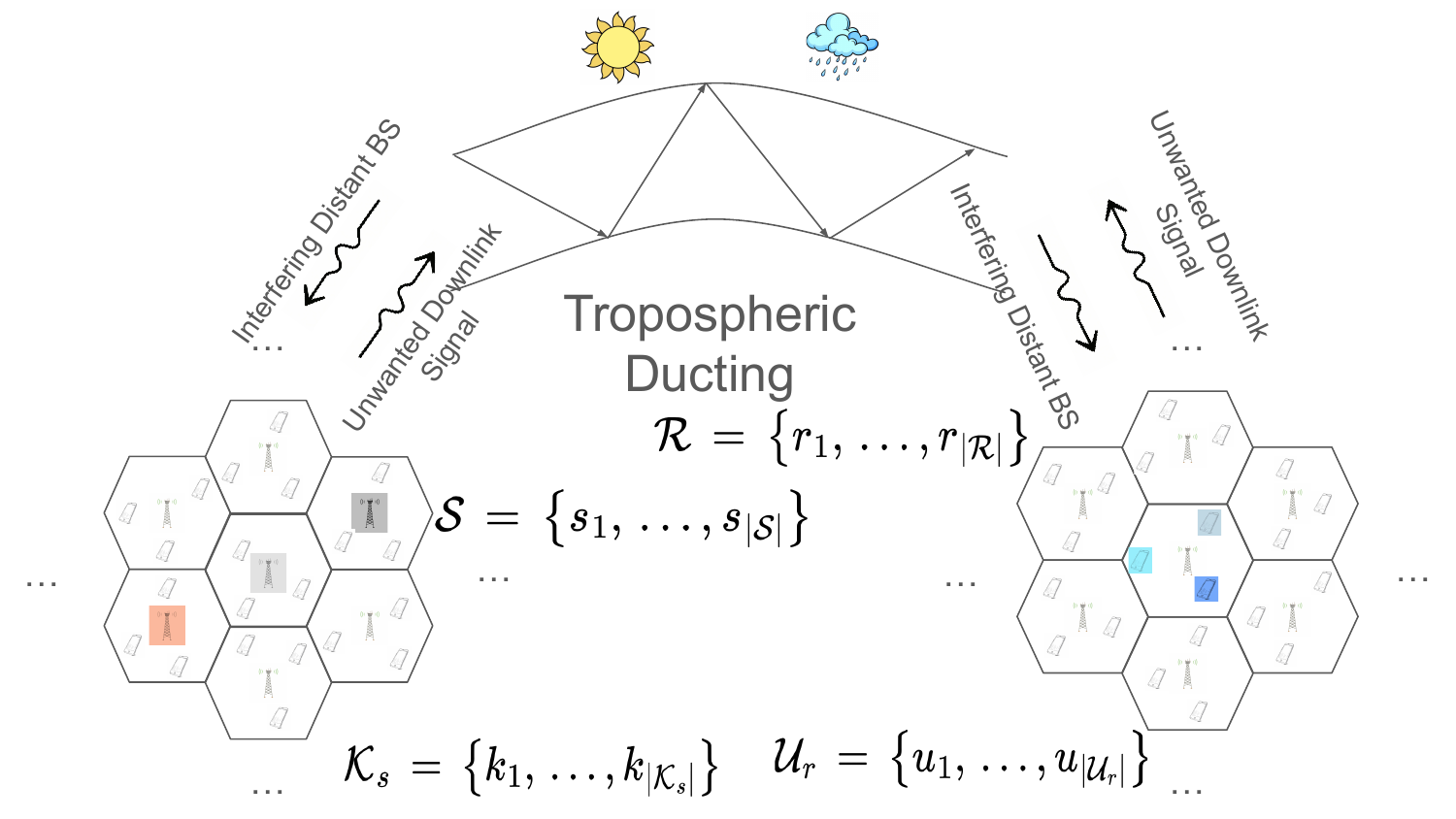}
		\vspace{-0.5em}
		\caption{Two cellular systems suffering from remote interference.}
		\label{fig:Geographic_View}
		\vspace{-1em}
	\end{figure}

	RI has unique characteristics making it fundamentally different from traditional interference sources such as inter-user-equipment (UE)-interference (IUI). Namely, RI has the following properties:
	\begin{itemize}
		\item It propagates hundreds of kilometers in the lower layer of atmosphere allowing it to affect the UEs of distant victim base stations (BSs). Typical distances are within~150~km in inland areas and can be farther than~300~km in coastal areas\cite{3GPP}.
		\item It affects the uplink (UL) transmission phase of victim cells due to the large propagation delay.
		\item It can be more powerful than the UL signals transmitted by UEs~\cite{power comparison}.
		\item It stays for long durations; e.g., minutes to hours.
	\end{itemize}
	RI is caused by several natural effects such as solar flares and coronal mass ejections~\cite{nature1}, precipitation and clouds~\cite{nature2}, lightning and atmospheric discharges~\cite{nature3}, and geomagnetic storms~\cite{nature4}. These natural effects create a tropospheric ducting layer that acts as a ``waveguide'' for signals transmitted by BSs located several kilometers away. Interfering BSs are called aggressor BSs, while the BSs affected by RI are called victim BSs. Interestingly, the created ``waveguide" is dominated by a line of sight (LoS) path with low path loss. In Figure~\ref{fig:Geographic_View}, we describe the phenomenon, showing a tropospheric duct layer that connects distant cellular systems.
	
	One important point to emphasize is that due to propagation delays, RI coming from the DL of the aggressor BSs, can affect both the CE and UL transmission phases of the victim's BSs. This effect is reciprocal in the sense that an aggressor BS can be a victim BS, since the tropospheric ducting layer allows for a two-way propagation.
	
	To deal with RI, several methods can be proposed~\cite{RI survey}:
	\begin{itemize}
		\item Time domain schemes: Increasing the guard period (GP) between the DL and the UL transmission phases (DL ends earlier and UL starts later)~\cite{RI Time}
		\item Frequency domain schemes: Assigning orthorgonal frequencies/subcarriers for victims and aggressors~\cite{RI Freq}. This will require an effective scheme to coordinate the BSs.
		\item Spatial domain schemes: Adjusting BS's antenna down-tilting, beam nulling, and mechanical/electrical designs to remove the atmospheric duct.
		\item Power domain schemes: Increases the UL power of UEs while decreasing aggressor BSs' DL power. This would result in a significant drawback, as it would reduce the battery life of UEs. As we will see in our results, our approach provides a better alternative, where UEs can enjoy better service without increasing their power consumption.
	\end{itemize}
	These passive schemes incur significant resource costs. For instance, extending the GP across coherence blocks over minutes to hours wastes substantial time. Similarly, assigning BSs to orthogonal frequency subsets wastes bandwidth, down-tilting antennas reduces coverage, and adjusting BS/UE transmit power either shortens UE battery life or degrades DL AR.
	
	In this paper, we study the impact of RI on CE quality and UL transmission. We analyze the CE, UL, and DL phases under three scenarios: (i) RI ignored, serving as a benchmark; (ii) a null-precoding scheme to suppress RI; and (iii) a fractional programming approach that optimizes aggressor BS precoders for more considerate DL transmissions. Results show that the proposed active method effectively mitigates RI, enabling high performance even in low transmit power regimes, while significantly improving CE accuracy and UE data rates, with RI nearly eliminated at high UL transmit powers.% is when we  and two precoding \& combining methods. %We have introduced a new method that balances between both sides of victims and aggressors.
	
	The rest of the paper is organized as follows. Section~\ref{section: system model} introduces the system model. Section~\ref{section: methods} presents the CE and UL phase under different scenarios. Section~\ref{section: results} provides the numerical results with key insights. Finally, Section~\ref{section: conclusion} summarizes the conclusion and the lessons learned.
	
	\textit{Notation:} 
	%italics to denote scalars, e.g., $r$, boldface to denote vectors, e.g., $\mathbf{h}$ and the calligraphic font to denote sets, e.g., $\mathcal{R}$. $\mathbb{C}^{M\times N}$ denotes the set of $M \times N$ matrices of complex numbers ($N=1$ denotes vectors). $\mathcal{CN}(\boldsymbol{\mu}, \mathbf{R})$ denotes the complex normal distribution with mean $\boldsymbol{\mu}$ and covariance matrix $\mathbf{R}$. $\mathbf{I}_M$ is the $M\times M$ identity matrix, and 
	$\mathcal{CN}(\boldsymbol{\mu}, \mathbf{R})$ denotes the complex normal distribution with mean $\boldsymbol{\mu}$ and covariance matrix $\mathbf{R}$. $\mathbb{E}[\cdot]$ denotes expectation; superscripts $(\cdot)^{\rm GP}$, $(\cdot)^{\rm ce}$, $(\cdot)^{\rm ul}$, and $(\cdot)^{\rm dl}$ denote parameters in GP, CE, UL, and DL phases, respectively; $(\cdot)^*$ and $(\cdot)^\star$ denote conjugate and optimum solution, respectively.
	%Using superscript $(\cdot)^{\rm T}$ and $(\cdot)^{\rm H}$ to denote transpose and hermitian, respectively. 
	
	\section{System Model}\label{section: system model}
	To study RI, we consider two distant cellular systems separated by a distance $l$, both operating in time division duplex (TDD) mode. We use $\mathcal{R}$ and $\mathcal{S}$ to represent the set of BSs in the two cellular systems, respectively. Each BS is identified by its index $1\leq r\leq |\mathcal{R}|$ and $1\leq s\leq |\mathcal{S}|$, respectively. There are $M$ antennas mounted as a linear array on each BS and one single antenna mounted on each UE. We use $\mathcal{U}_r$ in cell $r$, $r = 1,\dots,|\mathcal{R}|$, and $\mathcal{K}_s$ in cell $s$, $s = 1,\dots,|\mathcal{S}|$ to represent the set of UEs of each cell of two cellular systems, respectively.
	
\begin{figure}
		\centering
		\includegraphics[width=1\linewidth]{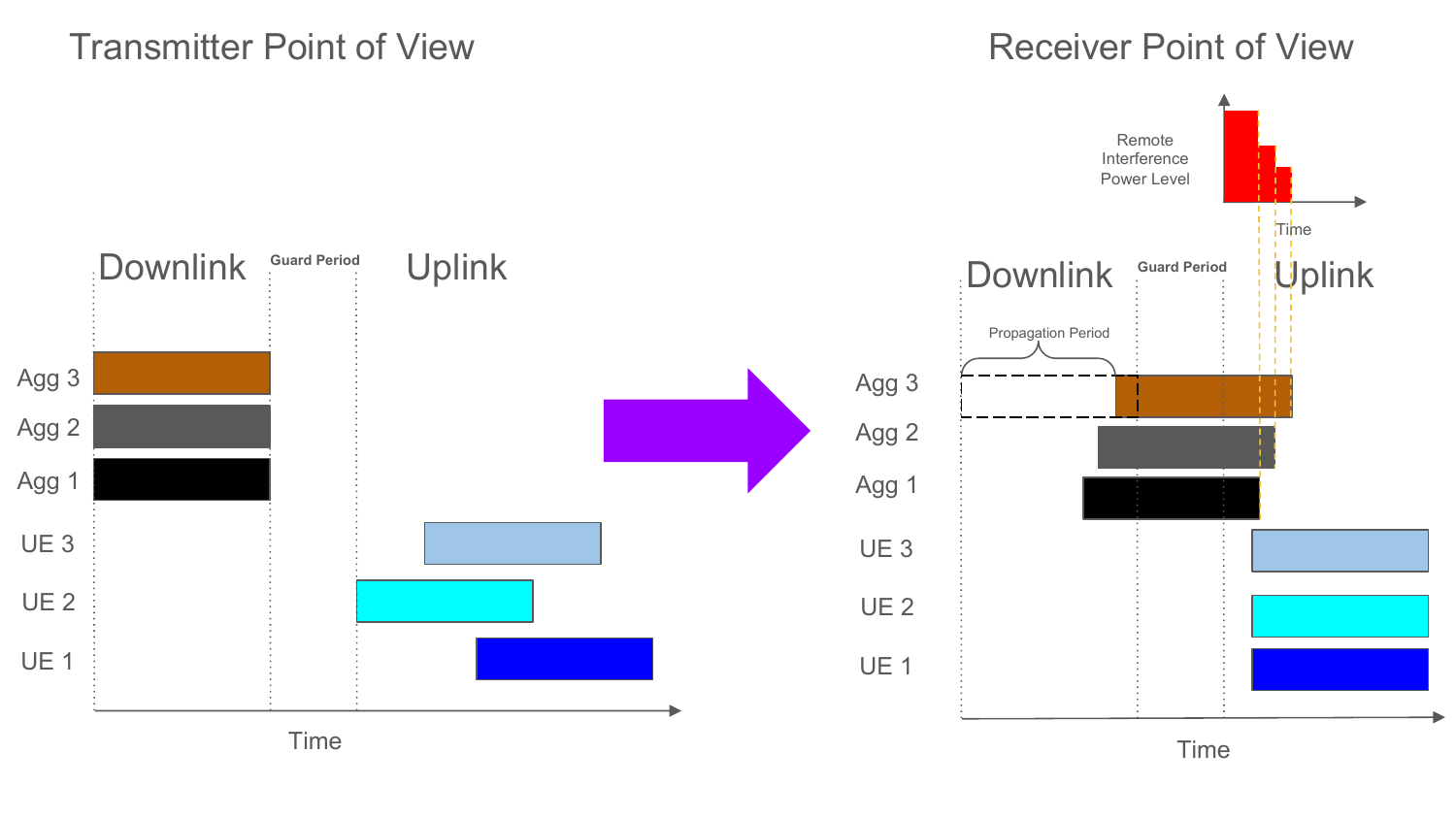}
		\vspace{-3em}
		\caption{Time View}
		\label{fig:Time_View}
		\vspace{-1.5em}
	\end{figure}

	Figure~\ref{fig:Time_View} illustrates how the DL transmissions of aggressor BSs (denoted as ``Agg'') can affect the CE and UL phases of UEs in victim cells (denoted as UE 1, UE 2, and UE3). The different colors of the rectangular blocks denote the signal sequences transmitted from corresponding transmitters highlighted with the same color in Figure~\ref{fig:Geographic_View}. 
	%UEs highlighted in blue, according to their location refer to their serving BS, they have been scheduled to start transmission sooner or later to guarantee their signal sequences will be received synchronously within the CE and UL transmission phases. On the aggressor side, as the DL transmission phase starts, all aggressor BSs start transmission at the same time with the same sequence length. 
	The GP is normally introduced between adjacent channel coherence blocks to avoid DL and UL transmission overlaps. However, due to the large distances between the aggressor and victim cells, propagation delay can easily result in reception delay of RI that is longer than the GP. This causes the RI to interfere with the pilot signals sent during the CE phase and the UE's data sent during the UL phase. Since the locations of the aggressor BSs' are different, RI from each aggressor ends at different times. As a result, the power level of RI at the victim cells exhibits a step-down pattern over time.
	
	We establish a restricted area with a radius of $\gamma$~meters around each BS, preventing UEs from being located too close to the BSs. This approach ensures reliable results by avoiding proximity effects and simulates a scenario where the APs are deployed at a height of $\gamma$~meters, serving UEs at ground level.
	
	We define the channel between BS $r$ and UE $u$ as $\boldsymbol{h}_{ru}\triangleq\sqrt{\beta_{ru}\psi_{ru}}\boldsymbol{g}_{ru}\in\mathbb{C}^{M\times 1}$, where $\beta_{ru}$, $\psi_{ru}$, and $\boldsymbol{g}_{ru}\sim\mathcal{CN}(\mathbf{0},\boldsymbol{I}_M)$ accounts for the effects of path-loss, shadowing, and the Rayleigh fading component, respectively. Besides, we define the duct channel between BS $r$ and $s$ as $\boldsymbol{H}_{rs}\in\mathbb{C}^{M\times M}$ which follows a Rician distribution. This channel is defined~as
	\begin{align}
		&\boldsymbol{H}_{rs} \triangleq \sqrt{\frac{KL}{K + 1}}\boldsymbol{q}_{r}^{\rm Rx}(\theta_r)\boldsymbol{q}_{s}^{\rm Tx}(\phi_s)^{\rm H} + \sqrt{\frac{L}{K + 1}}\boldsymbol{H}_{rs}^{\rm NLoS}\nonumber\\
		\mathrm{where~}
		&\boldsymbol{H}_{rs}^{\rm NLoS}\sim\mathcal{CN}\left(\boldsymbol{0}^{M\times M},M\boldsymbol{I}_M\right)\nonumber\\
		&\boldsymbol{q}(\theta) = \left[1, e^{j 2\pi \frac{d}{\lambda} \sin(\theta)}, \dots, e^{j 2\pi \frac{(M-1)d}{\lambda} \sin(\theta)}\right]^{\rm T}
		%	&\theta_r,~\phi_s\overset{\mathrm{iid}}{\sim}\mathrm{uni}\left[-\frac{\delta}{2},\frac{\delta}{2}\right]
	\end{align}
	Herein, $K$ is the Rician K-Factor, $L$ is the path-loss through the duct layer, $\boldsymbol{q}_{r}^{\rm Rx}(\theta_r)$ and $\boldsymbol{q}_{s}^{\rm Tx}(\phi_s)$ are the receiving and transmitting steering vectors, respectively. Also, $\theta_r$ and $\phi_s$ are the angle of arrival (AoA) of the victim BS $r$ and angle of departure (AoD) of the aggressor BS $s$, respectively. Both $\theta_r$ and $\phi_s$ are uniformly distributed $[-\delta/2,\delta/2]$, where $\delta$ is the maximum angular spread of AoD and AoA of RI. For simplicity, we assume it is fixed among all BSs.
	
	\section{Remote Interference Mitigation}\label{section: methods}
	In this section, we analyze the impact of RI on the CE phase and the UL transmission phase. We then design two combiners at the victim side which employs either maximum ratio combining (MRC) or minimum mean squared error (MMSE) based combiner. For the aggressor side, we develop two precoders that allow the aggressors to be more considerate in their DL transmissions. This is done through two precoder designs which are null precoding and fractional-programming (FP) based precoders. Finally, we analyze the impact of our schemes on CE accuracy, UL transmission at the victim side, and DL transmission at the aggressor side.
	
	\subsection{Analysis of Remote Interference}
	The signal received in the CE phase at BS~$r$, can be represented by matrix $\boldsymbol{Y}_{r}^{\rm ce}\in\mathbb{C}^{M\times\tau_{\rm p}}$ as,
	\begin{align}
		&\boldsymbol{Y}_{r}^{\rm ce} = \overbrace{\sqrt{p^{\rm ul}}\sum_{r'\in\mathcal{R}}\sum_{u\in\mathcal{U}_{r'}}\boldsymbol{h}_{ru}\boldsymbol{\phi}_{u}^{\rm T}}^{\mathrm{Local~Pilots}} + \overbrace{\sqrt{p^{\rm dl}}\sum_{s\in\mathcal{S}}\boldsymbol{H}_{rs}\boldsymbol{X}_{s}}^{\mathrm{RI}}  + \overbrace{\boldsymbol{Z}_r}^{\mathrm{Noise}} \nonumber\\
		&\mathrm{where~~}\boldsymbol{X}_{s} = \boldsymbol{W}_s\boldsymbol{S}_s ,
	\end{align}
	and $p^{\rm ul}$ and $p^{\rm dl}$ denote the UL and DL transmit power levels, respectively. The symbol $\boldsymbol{\phi}_{u}\in\mathbb{C}^{\tau_{\rm p}\times1}$ is the pilot sequence of UE $u$, $\boldsymbol{X}_s\in\mathbb{C}^{M\times\tau_{\rm p}}$ is the data sequence transmitted by BS $r$, $\tau_{\rm p}$ is the pilot length, and both $\boldsymbol{W}_s\in\mathbb{C}^{M\times|\mathcal{U}_s|}$ and $\boldsymbol{S}_s\in\mathbb{C}^{|\mathcal{U}_s|\times\tau_{\rm p}}$ are the precoder codebook and the signal matrix of the aggressor BS $s$, respectively. The noise matrix is represented though $\boldsymbol{Z}_r$, the entries of $\boldsymbol{Z}_r$ are independent and distributed as $\mathcal{CN}(0,\sigma^2)$.
	% $\boldsymbol{Z}_r\sim\mathcal{CN}\left(\boldsymbol{0}_M,\tau_{\rm p}\sigma^2\boldsymbol{I}_M\right)$, where $\boldsymbol{0}_M$ denotes an $M\times M$ matrix with all zero entries.
	
	We use match-filtering (MF) and linear (LMMSE) estimation for CE, where we use Discrete Fourier Transform (DFT)-based pilot sequences allowing for
	\begin{align}
		\boldsymbol{\phi}_{u}^{\rm H}\boldsymbol{\phi}_{u'} =
		\begin{cases}
			\tau_{\rm p} &{\text{if $u = u'$}}\\
			0 &{\text{if $u \neq u'$}}
		\end{cases}
	\end{align}
	To perform CE, we first use MF to filter out all non-copilot interference as
	\begin{align}
		\Breve{\boldsymbol{y}}_{ru} &= \frac{\boldsymbol{Y}_{r}^{\rm ce}\boldsymbol{\phi}_u^*}{\sqrt{p^{\rm ul}}}\nonumber\\
		&= \tau_{\rm p}\sum_{u\in\mathcal{C}_u}\boldsymbol{h}_{ru} + \sqrt{p^{\rm dl}}\sum_{s\in\mathcal{S}}\frac{\boldsymbol{H}_{rs}\boldsymbol{X}_{s}\boldsymbol{\phi}_u^*}{\sqrt{p^{\rm ul}}} + \frac{\boldsymbol{Z}_r\boldsymbol{\phi}_u^*}{\sqrt{p^{\rm ul}}}
	\end{align}
	Where $\mathcal{C}_u$ denotes the co-pilot UE set of the UE $u$. Then, we use LMMSE to suppress the remote interference and estimate the channel state information (CSI) between each BS~$r$ and its serving UE~$u$ as
	\begin{align}
		\hat{\boldsymbol{h}}_{ru} &= \boldsymbol{\Sigma}_{\boldsymbol{h}_{ru}\Breve{y}_{ru}}\boldsymbol{\Sigma}_{\Breve{y}_{ru}}^{-1}\Breve{\boldsymbol{y}}_{ru}\nonumber\\
		&= \tau_{\rm p}\beta_{ru}\psi_{ru}\Biggl[\left(\tau_{\rm p}^2\sum_{u\in\mathcal{C}_u}\beta_{ru}\psi_{ru} + \frac{\sigma^2\tau_{\rm p}}{p^{\rm ul}}\right)\boldsymbol{I}_M\nonumber\\
		&\hspace{-7.5mm}+ \frac{p^{\rm dl}}{p^{\rm ul}}\sum_{s\in\mathcal{S}}\tau_{\rm p}\left(\frac{KL}{K+1}\boldsymbol{q}_{r}^{\rm Rx}\boldsymbol{q}_{r}^{\rm Rx,H} + \frac{L}{K + 1}\boldsymbol{I}_M\right)\Biggl]^{-1}\breve{\boldsymbol{y}}_{ru}
	\end{align}
	According to LMMSE estimation, the estimated channel $\hat{\boldsymbol{h}}_{ru}$ is distributed according to $\mathcal{CN}\left(\boldsymbol{0},\boldsymbol{\Sigma}_{\hat{\boldsymbol{h}}_{ru}}\right)$, with the covariance matrix calculated as
	\allowdisplaybreaks
	\begin{align}
		\boldsymbol{\Sigma}_{\hat{\boldsymbol{h}}_{ru}} &\triangleq \boldsymbol{\Sigma}_{\boldsymbol{h}_{ru}\Breve{y}_{ru}}\boldsymbol{\Sigma}_{\Breve{y}_{ru}}^{-1}\boldsymbol{\Sigma}_{\Breve{\boldsymbol{y}}_{ru}\boldsymbol{h}_{ru}}\nonumber\\
		&= \Biggl[\left(\tau_{\rm p}^2\sum_{u\in\mathcal{C}_u}\beta_{ru}\psi_{ru} + \frac{\sigma^2\tau_{\rm p}}{p^{\rm ul}}  + \frac{|\mathcal{S}|L}{K + 1}\right)\boldsymbol{I}_M\nonumber\\
		\label{eq: est_h_cov}
		&\hspace{-7.5mm}+ \frac{p^{\rm dl}}{p^{\rm ul}}\sum_{s\in\mathcal{S}}\tau_{\rm p}\left(\frac{KL}{K+1}\boldsymbol{q}_{r}^{\rm Rx}\boldsymbol{q}_{r}^{\rm Rx,H}\right)\Biggl]^{-1}\tau_{\rm p}^2\beta_{ru}^2\psi_{ru}^2
	\end{align}
	The resulted channel estimation error is $\boldsymbol{e}_{ru} = \boldsymbol{h}_{ru} - \hat{\boldsymbol{h}}_{ru}$, and its distribution is $\mathcal{CN}\left(\boldsymbol{0},\boldsymbol{\Sigma}_{\boldsymbol{e}_{ru}}\right)$, where $\boldsymbol{\Sigma}_{\boldsymbol{e}_{ru}} = \boldsymbol{\Sigma}_{\boldsymbol{h}_{ru}} - \boldsymbol{\Sigma}_{\hat{\boldsymbol{h}}_{ru}}$. Due to the RI in \eqref{eq: est_h_cov}, the third scalar for $\boldsymbol{I}_M$ and the second summation term of $s\in\mathcal{S}$, they both decrease the scale of $\boldsymbol{\Sigma}_{\hat{\boldsymbol{h}}_{ru}}$, which also increases the scale of $\boldsymbol{\Sigma}_{\boldsymbol{e}_{ru}}$.
	
	After estimating the channels between each BS and its UEs, the UL transmission result in the following received UL signal vector at each BS~$r$
	\begin{align}
		\boldsymbol{y}_r^{\rm ul} = \sqrt{p^{\rm ul}}\sum_{r'\in\mathcal{R}}\sum_{u\in\mathcal{U}_{r'}}\boldsymbol{h}_{ru}s_u + \sqrt{p^{\rm dl}}\sum_{s\in\mathcal{S}}\boldsymbol{H}_{rs}\boldsymbol{x}_s + \boldsymbol{z}_r ,
	\end{align}
	where $\boldsymbol{x}_s = \boldsymbol{W}_s\boldsymbol{s}_s$ with $\boldsymbol{s}_s$ being the signal vector transmitted by BS $s$ (an aggressor BS). The noise vector $\boldsymbol{z}_r\sim\mathcal{CN}\left(\boldsymbol{0}^{M\times1},\sigma^2\boldsymbol{I}_M\right)$, where $\boldsymbol{0}^{M\times1}$ denotes a column vector with length of $M$ and all zero entries.
	%Then, the instantaneous UL achievable rate (AR) at BS $r$ for UE $u$ can be expressed as \eqref{eq: uplink AR suffer}.
	%\begin{figure*}[b]
	%	\begin{align}\label{eq: uplink AR suffer}
		%		A_{ru}^{\rm ul} = \log_2\left(1 + \epsilon_{ru}\right), ~~\epsilon_{ru} = \frac{p^{\rm ul}\left|\boldsymbol{c}_{ru}^{\rm H}\hat{\boldsymbol{h}}_{ru}\right|^2}{p^{\rm ul}\left|\boldsymbol{c}_{ru}^{\rm H}\boldsymbol{e}_{ru}\right|^2 + p^{\rm ul}\sum_{u'\neq u}\left|\boldsymbol{c}_{ru}^{\rm H}\boldsymbol{h}_{ru'}\right|^2 + \frac{p^{\rm dl}}{M}\sum_{s\in\mathcal{S}}\left|\boldsymbol{c}_{ru}^{\rm H}\boldsymbol{H}_{rs}\right|^2 + \sigma^2\norm{\boldsymbol{c}_{ru}}^2}
		%	\end{align}
	%\end{figure*}

	For the the UL data transmission phase, each BS needs to design a combiner. For this purpose, we employ either an MRC or MMSE combiner defined as
	\begin{align}
		\boldsymbol{c}_{ru}^{\rm MRC} &= \frac{\hat{\boldsymbol{h}}_{ru}}{\norm{\hat{\boldsymbol{h}}_{ru}}}\\
		\boldsymbol{c}_{ru}^{\rm MMSE} &= \boldsymbol{\Sigma}_{\boldsymbol{y}_{r}}^{-1}\boldsymbol{\Sigma}_{\boldsymbol{y}_rs_u}=\\
	&\hspace{-10mm} \Biggl\{p^{\rm ul}\left[\sum_{u\in\mathcal{U}_r}\left(\hat{\boldsymbol{h}}_{ru}\hat{\boldsymbol{h}}_{ru}^{\rm H} + \boldsymbol{\Sigma}_{\mathrm{e},ru}\right) + \sum_{u''\in\mathcal{U}_r'}\beta_{ru''}\psi_{ru''}\boldsymbol{I}_M\right] + \nonumber\\
		&\hspace{-10mm}\sum_{s\in\mathcal{S}}\frac{p^{\rm dl}KL}{K+1}\boldsymbol{q}_{rs}^{\rm Rx}\boldsymbol{q}_{rs}^{\rm Rx,H} + \frac{p^{\rm dl}|\mathcal{S}|L}{K + 1}\boldsymbol{I}_M + \sigma^2\boldsymbol{I}_M\Biggl\}^{-1}\sqrt{p^{\rm ul}}\hat{\boldsymbol{h}}_{ru}\nonumber
	\end{align}
	where $\mathcal{U}_r' = \bigcup_{r'\in\mathcal{R},r'\neq r}\mathcal{U}_{r'}$.
	
	However, we can see that in both LMMSE channel estimation and MMSE combiner design, we need the steering vector of AoA of the RI at BS~$r$, $\phi_r$. To do so, we propose using the root-MUlti SIgnal Classification (MUSIC) algorithm.

	\subsection{Root-MUSIC}
	The sampled matrix $\boldsymbol{Y}_r^{\rm GP}$ is obtained by collecting the last samples of the RI signals received during the GP across $P$ coherence time blocks,
	\begin{align}
		\boldsymbol{Y}_r^{\rm GP} &= \left[\boldsymbol{y}_{r,1}^{\rm GP},\dots,\boldsymbol{y}_{r,P}^{\rm GP}\right]
		\\
		\boldsymbol{y}_{r}^{\rm GP} &= \sqrt{p^{\rm dl}}\sum_{s\in\mathcal{S}}\boldsymbol{H}_{rs}\boldsymbol{x}_s + \boldsymbol{z}_r
	\end{align}
	Then, eigenvalue decomposition can be performed on the sampled covariance matrix $\boldsymbol{\Sigma}_{\boldsymbol{y}_{r}^{\rm GP}}^{\rm SP}$ to find a matrix $\boldsymbol{Q}_r^{\rm n}$ that contains the basis of the noise only subspace,
	\begin{align}
		\boldsymbol{\Sigma}_{\boldsymbol{y}_{r}^{\rm GP}}^{\rm SP} &= \frac{1}{P}\boldsymbol{Y}_r^{\rm GP}\boldsymbol{Y}_r^{\rm GP,H}\\
		\boldsymbol{Q}_r\boldsymbol{\Lambda}_r\boldsymbol{Q}_r^{\rm H} &= \boldsymbol{\Sigma}_{\boldsymbol{y}_{r}^{\rm GP}}^{\rm SP}\nonumber\\
		\boldsymbol{Q}_r &= \left[\boldsymbol{q}_{r,1},\dots,\boldsymbol{q}_{r,|\mathcal{S}|},\boldsymbol{Q}_r^{\rm n}\right]
	\end{align}
	If we denote $z(\theta_r) = e^{-j 2\pi \frac{d}{\lambda} \sin(\theta_r)}$, where $d$ and $\lambda$ are the distance between adjacent antennas and wavelength, respectively. Then, the function solving for roots according to \cite{rootMUSIC},
	\begin{gather}
		\sum_{l = 0}^{2M-2}z^lC_l = 0\nonumber\\
		\mathrm{where~~}\boldsymbol{C} = \boldsymbol{Q}_r^{\rm n}\boldsymbol{Q}_r^{\rm n,H},~C_l = \sum_{n - m = l}\boldsymbol{C}_{mn}
	\end{gather}
	This is the root-MUSIC that estimates AoA for precoder design to mitigate RI.
%	root-MUSIC algorithm which will help estimate the AoA that will be used later to design the precoders that mitigate RI.
%	\vspace{-17.5mm}
	\subsection{Null Precoding}
	We have seen that the troposhereic ducting layer acts as a waveguide for the RI. Notice that the RI channel follows a Rician distribution with a strong K-Factor, hence, if we can block its LoS component, we can greatly mitigate it. Based on this idea, we design a null precoder, which is designed using the null space of the steering vector of the AoD, $\boldsymbol{a}(\phi_s)$, i.e.,
	\begin{align}
		\boldsymbol{W}_s^{\rm null} = \mathrm{null}\left[\boldsymbol{a}(\phi_s)\right]
	\end{align}
	If the LoS component is blocked, the signal model and the LMMSE channel estimator can be reformulated as,
	\begin{align}
		&\hspace{-7mm}\boldsymbol{Y}_{r}^{\rm ce} = \sqrt{p^{\rm ul}}\sum_{\substack{r\in\mathcal{R}\\u\in\mathcal{U}_r}}\boldsymbol{h}_{ru}\boldsymbol{\phi}_{u}^{\rm T} + \sqrt{\frac{p^{\rm dl}L}{K + 1}}\sum_{s\in\mathcal{S}}\boldsymbol{H}_{rs}^{\rm NLoS}\boldsymbol{X}_{s}^{\rm null} + \boldsymbol{Z}_r\nonumber\\
		&\hspace{-7mm}\mathrm{where~~}\boldsymbol{X}_{s}^{\rm null} = \boldsymbol{W}_s^{\rm null}\boldsymbol{S}_s\\
		\hat{\boldsymbol{h}}_{ru}^{\rm null} &= \tau_{\rm p}\beta_{ru}\psi_{ru}\Biggl[\left(\tau_{\rm p}^2\sum_{u\in\mathcal{C}_u}\beta_{ru}\psi_{ru} + \frac{\sigma^2\tau_{\rm p}}{p^{\rm ul}}\right)\boldsymbol{I}_M\nonumber\\
		&\hspace{5mm}+ \frac{p^{\rm ul}\tau_{\rm p}|\mathcal{S}|(|\mathcal{U}_s| - 1)L}{p^{\rm dl}(K + 1)}\boldsymbol{I}_M\Biggl]^{-1}\breve{\boldsymbol{y}}_{ru}\nonumber\\
		&= \frac{\tau_{\rm p}\beta_{ru}\psi_{ru}\breve{\boldsymbol{y}}_{ru}}{\tau_{\rm p}^2\sum_{u\in\mathcal{C}_u}\beta_{ru}\psi_{ru} + \frac{\sigma^2\tau_{\rm p}}{p^{\rm ul}} + \frac{p^{\rm ul}\tau_{\rm p}|\mathcal{S}|(|\mathcal{U}_s| - 1)L}{p^{\rm dl}(K + 1)}}
	\end{align}
	Then, the new covariance matrix for $\hat{\boldsymbol{h}}_{ru}^{\rm null}$ and $\boldsymbol{e}_{ru}$ becomes,
	\begin{align}
		\boldsymbol{\Sigma}_{\hat{\boldsymbol{h}}_{ru}^{\rm null}} &= \frac{\tau_{\rm p}^2\beta_{ru}^2\psi_{ru}^2}{\tau_{\rm p}^2\sum_{u\in\mathcal{C}_u}\beta_{ru}\psi_{ru} + \frac{\sigma^2\tau_{\rm p}}{p^{\rm ul}} + \frac{p^{\rm ul}\tau_{\rm p}|\mathcal{S}|(|\mathcal{U}_s| - 1)L}{p^{\rm dl}(K + 1)}}\boldsymbol{I}_M\\
		\boldsymbol{\Sigma}_{\boldsymbol{e}_{ru}}^{\rm null} &= \boldsymbol{\Sigma}_{\boldsymbol{h}_{ru}} - \boldsymbol{\Sigma}_{\hat{\boldsymbol{h}}_{ru}^{\rm null}} = \epsilon_{ru}\boldsymbol{I}_M
	\end{align}
	Given this, the MMSE combiner $\boldsymbol{c}_{ru,\rm MMSE}^{\rm null}$ needs to be re-designed as,
	\allowdisplaybreaks
	\begin{align}
		\boldsymbol{c}_{ru,\rm MMSE}^{\rm null} &= \Biggl[p^{\rm ul}\left(\hat{\boldsymbol{h}}_{ru}\hat{\boldsymbol{h}}_{ru}^{\rm H} + \boldsymbol{\Sigma}_{\mathrm{e},ru}^{\rm null}\right)\\
		&\hspace{-10mm}+ \left(p^{\rm ul}\sum_{u'\neq u}\beta_{ru'}\psi_{ru'} + \frac{p^{\rm dl}|\mathcal{S}|L}{K + 1} + \sigma^2\right)\boldsymbol{I}_M\Biggl]^{-1}\sqrt{p^{\rm ul}}\hat{\boldsymbol{h}}_{ru}\nonumber
		%	\mathrm{where~~}\alpha_{ru} &= \left(p^{\rm ul}\sum_{u'\neq u}\beta_{ru'}\psi_{ru'} + \sigma^2\right) + \frac{p^{\rm dl}ML|\mathcal{S}|}{|\mathcal{U}_s|\left(K + 1\right)}\nonumber\\
		%	&\hspace{-20mm}+ p^{\rm ul}\left[\beta_{ru}\psi_{ru} - \tau_{\rm p}^2\beta_{ru}^2\psi_{ru}^2\left(\tau_{\rm p}^2\sum_{u'\in\mathcal{C}_u}\beta_{ru'}\psi_{ru'} + \frac{p^{\rm dl}}{p^{\rm ul}}\frac{ML|\mathcal{S}|}{|\mathcal{U}_s|(K + 1)} + \frac{\sigma^2\tau_{\rm p}}{p^{\rm ul}}\right)^{-1}\right]
	\end{align}
	However, we notice that the null precoder is designed without considering the local CSI between the aggressor BS and its served UEs. Although most of the RI could be blocked by the null precoder, the local DL service of the aggressor BS would be poor and its DL achievable rate (AR) would be low as well.
	
	Therefore, our precoder should be designed to balance between the DL AR at the aggressor side and also the UL AR at the victim side, which we will do next.
	
	\subsection{Fractional Programming-based Precoding}
	To solve the issue of poor DL services for aggressor BSs, we design a new objective function that is the summation of the victim UL AR and the aggressor DL AR shown as~\eqref{Rul} and~\eqref{Rul}, respectively.
	\begin{figure*}[t]
	\begin{align}
        \label{Rul}
		R^{\rm ul}_{r} &= \sum_{u\in\mathcal{U}_{r}}\log\left(1 + \frac{p^{\rm ul}|\hat{\boldsymbol{h}}_{ru}^{\rm H}\boldsymbol{c}_{ru}|^2}{p^{\rm ul}\left(\underset{u\in\mathcal{U}_r}{\sum}\epsilon_{ru} + \underset{u''\in\mathcal{U}_{r}'}{\sum}\beta_{ru''}\psi_{ru''}\right)||\boldsymbol{c}_{ru}||^2 + p^{\rm ul}\underset{\substack{u'\in\mathcal{U}_{r}\\u'\neq u}}{\sum}|\hat{\boldsymbol{h}}_{ru'}^{\rm H}\boldsymbol{c}_{ru}|^2 + p^{\rm dl}\underset{s\in\mathcal{S}}{\sum}||\boldsymbol{c}_{ru}^{\rm H}\boldsymbol{H}_{rs}\boldsymbol{W}_{s}||^2 + \sigma^2}\right)\\
		% &= \sum_{u\in\mathcal{U}_{r}}\log\left(1 + p^{\rm ul}\boldsymbol{h}_{ru}^{\rm H}\left(p^{\rm ul}\sum_{k\in\{\mathcal{U}_{r}\backslash u\}}\boldsymbol{h}_{rk}\boldsymbol{h}_{rk}^{\rm H} + p^{\rm dl}\sum_{s\in\mathcal{R}'}\boldsymbol{H}_{rs}\boldsymbol{W}_{s}\boldsymbol{W}_{s}^{\rm H}\boldsymbol{H}_{rs}^{\rm H} + \sigma^2\boldsymbol{I}_M\right)^{-1}\boldsymbol{h}_{ru}\right)
        \label{Rdl}
            R^{\rm dl}_{s} &= \sum_{k\in\mathcal{K}_{s}}\log\left(1 + \frac{p^{\rm dl}|\boldsymbol{w}_{sk}^{\rm H}\hat{\boldsymbol{h}}_{sk}|^2}{p^{\rm dl}\left(\underset{k\in\mathcal{K}_s}{\sum}\epsilon_{sk} + \underset{k''\in\mathcal{K}_s'}{\sum}\beta_{sk''}\psi_{sk''}\right)||\boldsymbol{w}_{sk}||^2 + p^{\rm dl}\underset{\substack{k'\in\mathcal{K}_{s}\\k'\neq k}}{\sum}|\hat{\boldsymbol{h}}_{sk'}^{\rm H}\boldsymbol{w}_{sk}|^2 + \sigma^2}\right)
	\end{align}
		\hrule
	\end{figure*}
	We then formulate the following optimization problem that aims to design the precoder $\boldsymbol{W}_{s}$ at the aggressor BS that mitigates RI toward the victim BSs $\mathcal{R}$, while still considering the DL data rates for its UEs:
	\begin{subequations}\label{eq:optm_prob}
		\begin{align}
			\max_{\boldsymbol{W}_{s}} \sum_{r\in\mathcal{R}}R^{\rm ul}_{r} + \sum_{s\in\mathcal{S}}R^{\rm dl}_{s}
			\\
			\mathrm{s.t.} \sum_{k = 1}^{|\mathcal{K}_{s}|}||\boldsymbol{w}_{sk}||^2 = 1
		\end{align}
	\end{subequations}
	
	The problem in~\eqref{eq:optm_prob} is non-convex and hard to solve, however, it can be reformulated as a fractional program. The quadratic transform introduced in~\cite{FP} provides a well-defined method to solve fractional programs by introducing auxiliary variables $\gamma_u$, $\gamma_k$, $y_u$, and $y_k$ that allow obtaining a local optimum for~\eqref{eq:optm_prob}. Using the quadratic transform, we can convert the objective function to \eqref{eq: final},
	\begin{figure*}[b]
		\hrule
		\begin{subequations}
			\begin{align}
				&\max_{Y_u,Y_{k},\Gamma_r,\Gamma_{s},\boldsymbol{W}_{s}} \sum_{r\in\mathcal{R}}\sum_{u\in\mathcal{U}_{r}}\left\{\log\left(1 + \gamma_u\right) - \gamma_u + 2\Re\left(y_u^*\sqrt{(1+\gamma_u)p^{\rm ul}}\boldsymbol{h}_{ru}^{\rm H}\boldsymbol{c}_{ru}\right) - |y_u|^2\left(A_{u} + B_{u}\right)\right\}\nonumber\\
				\label{eq: final}
				&\hspace{15mm}+ \sum_{s\in\mathcal{S}}\sum_{k\in\mathcal{K}_{s}}\left[\log(1 + \gamma_{k}) - \gamma_{k} + 2\Re\left(y_{k}^*\sqrt{(1 + \gamma_{k})p^{\rm dl}}\boldsymbol{h}_{sk}^{\rm H}\boldsymbol{w}_{sk}\right) - |y_{k}|^2\left(C_{k} + D_{k}\right)\right]~~\mathrm{s.t.}\sum_{k = 1}^{|\mathcal{K}_{s}|}||\boldsymbol{w}_{sk}||^2 = 1
				\\
				&\mathrm{s.t.~}
				A_{u} = p^{\rm ul}\left|\hat{\boldsymbol{h}}_{ru}^{\rm H}\boldsymbol{c}_{ru}\right|^2, C_{k} = p^{\rm dl}\left|\hat{\boldsymbol{h}}_{sk}^{\rm H}\boldsymbol{w}_{sk}\right|^2, D_{k}=p^{\rm dl}\left(\underset{k\in\mathcal{K}_s}{\sum}\epsilon_{sk} + \underset{k''\in\mathcal{K}_s'}{\sum}\beta_{sk''}\psi_{sk''}\right)||\boldsymbol{w}_{sk}||^2 + p^{\rm dl}\underset{\substack{k'\in\mathcal{K}_{s}\\k'\neq k}}{\sum}|\hat{\boldsymbol{h}}_{sk'}^{\rm H}\boldsymbol{w}_{sk}|^2 + \sigma^2\nonumber\\
                &B_{u} = p^{\rm ul}\left(\underset{u\in\mathcal{U}_r}{\sum}\epsilon_{ru} + \underset{u''\in\mathcal{U}_{r}'}{\sum}\beta_{ru''}\psi_{ru''}\right)||\boldsymbol{c}_{ru}||^2 + p^{\rm ul}\underset{\substack{u'\in\mathcal{U}_{r}\\u'\neq u}}{\sum}|\hat{\boldsymbol{h}}_{ru'}^{\rm H}\boldsymbol{c}_{ru}|^2 + p^{\rm dl}\underset{s\in\mathcal{S}}{\sum}||\boldsymbol{c}_{ru}^{\rm H}\boldsymbol{H}_{rs}\boldsymbol{W}_{s}||^2 + \sigma^2
				\label{eq: aux update}
			\end{align}
		\end{subequations}
		\hrule
		\begin{align}
			&\gamma_u^\star = \frac{A_{u}}{B_{u}},
			\ \ \ \ \ 
			\gamma_k^\star = \frac{C_{k}}{D_{k}},
			\ \ \ \ \ 
			y_u^\star = \sqrt{(1 + \gamma_u)p^{\rm ul}}\hat{\boldsymbol{h}}_{ru}^{\rm H}\boldsymbol{c}_{ru}/(A_{u} + B_{u}),
			\ \ \ \ \ 
			y_k^\star = \sqrt{\left(1 + \gamma_k\right)p^{\rm dl}}\boldsymbol{w}_{sk}^{\rm H}\boldsymbol{h}_{sk}/\left(C_k + D_k\right)\\
			\label{eq: w_sk closed form}
			&\boldsymbol{w}_{sk} = \left\{\frac{p^{\rm dl}\norm{\boldsymbol{c}_{ru}}^2}{M}\left(\frac{KL}{K + 1}\boldsymbol{q}_s^{\rm Tx}\boldsymbol{q}_s^{\rm Tx,H} + \frac{LM}{K + 1}\boldsymbol{I}_M\right)\left(\sum_{u\in\mathcal{U}_r}|y_u|^2\right) + p^{\rm dl}\sum_{k\in\mathcal{K}_{s}}|y_{k}|^2\boldsymbol{h}_{sk}\boldsymbol{h}_{sk}^{\rm H} + \lambda_{s}\boldsymbol{I}_M\right\}^{-1} y_k\sqrt{(1+\gamma_{k})p^{\rm dl}}\boldsymbol{h}_{sk}
		\end{align}
		\vspace{-2em}
	\end{figure*}
	where $Y_u = \{\dots,y_{u},\dots\},~u\in\mathcal{U}_{r}$ and $Y_{k} = \{\dots,y_{k},\dots\},~k\in\mathcal{K}_{s}$. By fixing all the optimizing variables except one at-a-time, we can derive the first-order optimality condition of the objective function with respect to that variable. This allows us to obtain a closed-form expression or update rule for the optimal value of that variable.% Then, we need to determine the optimal formula for $y^\star$.
	
	The final step is to find the precoders $\boldsymbol{w}_{sk}$, by taking the Lagrangian dual of \eqref{eq: final},
	\begin{align}
		\mathcal{L} &= f_o + \sum_{s\in\mathcal{S}}\lambda_{s}\left(1 - \sum_{k\in\mathcal{K}_{s}}||\boldsymbol{w}_{sk}||^2\right)
		%	\frac{\partial \mathcal{L}}{\partial \boldsymbol{w}_{sk}} &= \sum_{u\in\mathcal{U}_r}\left(-|y_u|^2\frac{\partial B_u(\boldsymbol{w}_{sk})}{\partial\boldsymbol{w}_{sk}}\right) + 2y_k\sqrt{(1+\gamma_{k})p^{\rm dl}}\boldsymbol{h}_{sk} - |y_{k}|^2\frac{\partial C_k}{\partial \boldsymbol{w}_{sk}}\nonumber\\
		%	&\hspace{5mm} - \sum_{k'\in\{\mathcal{K}_{s}\backslash k\}}|y_{k'}|^2\frac{\partial D_{k'}}{\partial\boldsymbol{w}_{sk}} - 2\lambda_{s}\boldsymbol{w}_{sk}\nonumber\\
		%	&= \sum_{u\in\mathcal{U}_r}-|y_u|^2 p^{\rm dl}\left(\frac{2KL}{K+1}\boldsymbol{u}_{\rm Tx}(\phi)\boldsymbol{u}_{\rm Tx}(\phi)^{\rm H}\boldsymbol{w}_{sk} + \frac{2LM}{K+1}\boldsymbol{w}_{sk}\right)  + 2y_k\sqrt{(1+\gamma_{k})p^{\rm dl}}\boldsymbol{h}_{sk}\nonumber\\
		%	&\hspace{5mm} - 2p^{\rm dl}|y_{k}|^2\boldsymbol{h}_{sk}\boldsymbol{h}_{sk}^{\rm H}\boldsymbol{w}_{sk} - 2p^{\rm dl}\left(\sum_{k'\in\{\mathcal{K}_{s}\backslash k\}}|y_{k'}|^2\boldsymbol{h}_{sk'}\boldsymbol{h}_{sk'}^{\rm H}\right)\boldsymbol{w}_{sk} - 2\lambda_{s}\boldsymbol{w}_{sk}
	\end{align}
	Here, $f_o$ is the objective function of \eqref{eq: final}. Then, the closed-form solution for $\boldsymbol{w}_{sk}$ is given in \eqref{eq: w_sk closed form}, leading to Algorithm~\ref{alg:seq}.
%	Here, $f_o$ is the objective function of \eqref{eq: final}. Then, the closed form expression for $\boldsymbol{w}_{sk}$ can be shown as \eqref{eq: w_sk closed form}. Therefore, we can formulate the complete FP algorithm as shown in Algorithm~\ref{alg:seq} below.
	\begin{algorithm}[H]
		\begin{algorithmic}[1]
			\STATE Initialize $\boldsymbol{W}_s$ and $\eta$, set $i = 1$
			\WHILE{$|f_{\rm o}^{~i + 1} - f_{\rm o}^{~i}| > \eta f_{\rm o}^{~i}$}
			\STATE Update $\gamma_u$, $\gamma_k$, $y_u$, and $y_k$ according to \eqref{eq: aux update}
			\STATE Determine $\lambda_s$ using Binary Section Search (BSS) and update $\boldsymbol{W}_s$ according to \eqref{eq: w_sk closed form}
			\STATE i = i + 1
			\ENDWHILE
		\end{algorithmic}
		\caption{pseudocode for FP}
		\label{alg:seq}
	\end{algorithm}
	
	Under the FP precoding method, we use the same MMSE combiner design as null precoding method, since frequent information exchange between the BSs is not favored, and $\boldsymbol{W}_s$ designed by FP does not have a closed form expression. Then, the only thing that victim BS can do is hope that the aggressors' precoders has blocked most of RI, e.g., blocked the LoS component. Therefore, the error covariance matrix using for other designs is the same as $\boldsymbol{\Sigma}_{\boldsymbol{e}_{ru}}^{\rm null}$.
	
	\section{Numerical Results}\label{section: results}
	We set the distance between the two cellular systems to $l = 86~\rm km$, which is a common RI distance according to~\cite{3GPP}. We simulate with $50$~BSs on each side, $|\mathcal{R}| = |\mathcal{S}| = 50$, with $M = 64$ antennas at each BS. Each cell is a hexagon with one BS, and its side length is $250 \rm m$. UEs are uniformly distributed inside the cell with a number of $|\mathcal{U}_r| = |\mathcal{U}_s| = 7, \forall r, s$. The restricted area radius is $\gamma = 20 \rm m$. The pilot length $\tau_{\rm p} = 32$. We use the Walfisch-Ikegami path loss model whihc sets $\beta_{ru} = 10^{-11.2427}d_{ru}^{-3.8}$, where $d_{ru}$ denotes the distance between UE $u$ and BS $r$ in km; the shadowing modeled as, $\psi_{ru, \rm dB}\sim \mathcal{N}(0,\sigma^2_{\rm dB}),\:\sigma_{\rm dB} = 4$. The noise power is $\sigma^2 = 1\times 10^{-14}\rm W$. For simplicity, we assume there are no phase/timing synchronization errors among the antennas at each AP. The UL power, $p^{\rm ul}$, varies from $-30$ to $40$~dBm, and the DL power, $p^{\rm dl}$, is fixed at $40$~dBm. The K-Factor of the RI channel is $K = 1000$. The large-scale-fading-constant (LSFC) of the RI channel, $L$, is calculated according to \cite[Sec. 4.4]{itu}.

\begin{figure}[t]
		\centering
		\includegraphics[width=1\linewidth]{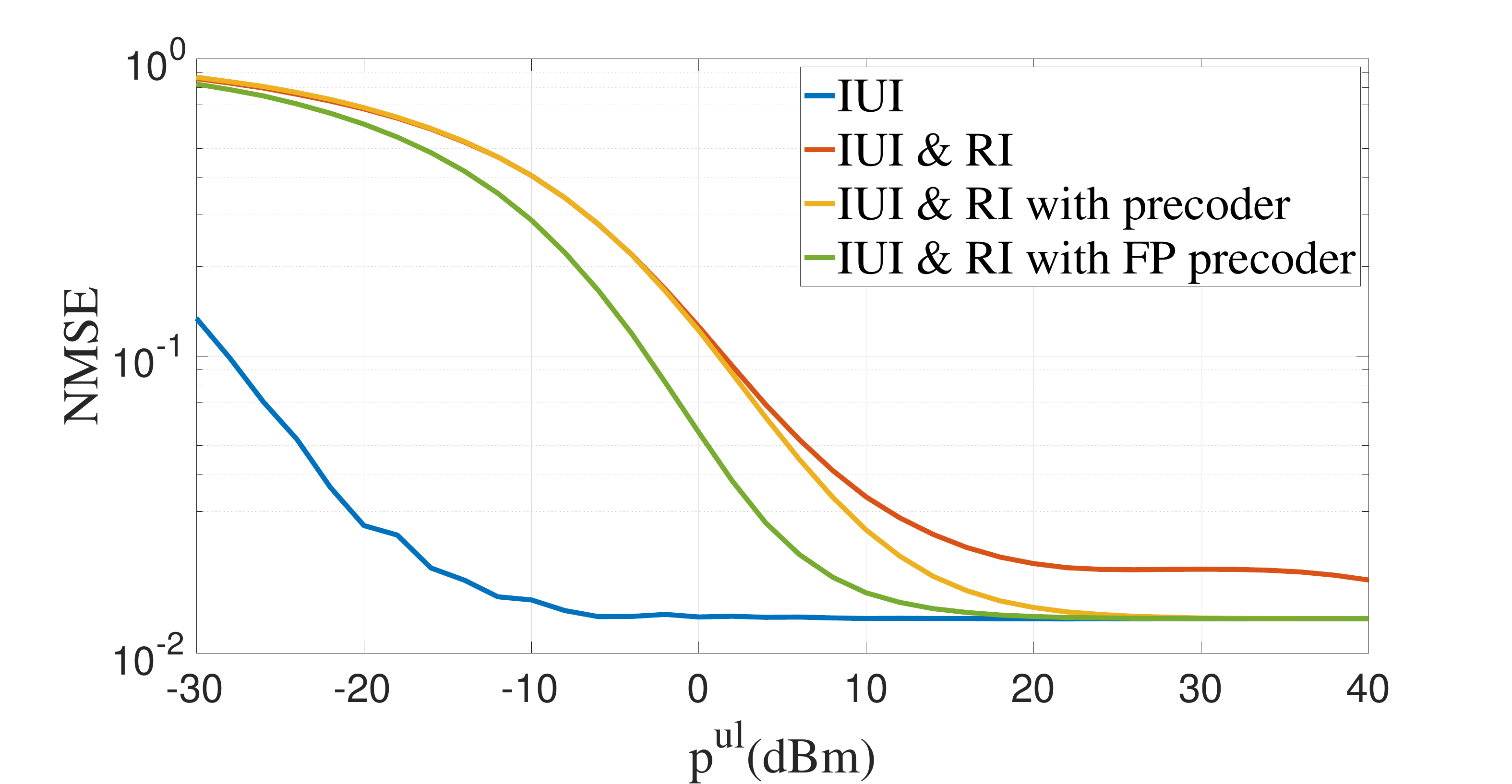}
		\vspace{-1.5em}
		\caption{NMSE vs Transmit Power Comparison in CE phase}
		\label{fig: NMSE Comparison}
		\vspace{-1.5em}
	\end{figure}

    We evaluate the normalized-mean-squared-error (NMSE), defined as $\mathrm{NMSE} = ||\boldsymbol{h}_{ru}-\hat{\boldsymbol{h}}_{ru}||^2/||\boldsymbol{h}_{ru}||^2$, against uplink power $p^{\rm ul}{|\rm dBm}$. As shown in Figure~\ref{fig: NMSE Comparison}, the blue curve (without RI) represents the optimum case and provides a lower bound. In contrast, the red curve (ignoring RI) performs worst, with NMSE between 0.04 and 0.05 at $p^{\rm ul} = 14~\rm dBm$. The purple curve converges to approximately $\mathrm{NMSE} \approx 0.015$. FP precoding outperforms null precoding (yellow curve), as null precoding only suppresses the LoS component of RI, while FP precoding mitigates both LoS and NLoS parts. At $p^{\rm ul} = 14~\rm dBm$, the yellow curve yields $\mathrm{NMSE} \approx 0.017$. Additionally, the red curve exhibits two slope changes, reflecting the dual nature of RI from independent LoS and NLoS sources.
	
	% We first compare the normalized-mean-squared-error (NMSE), as a function of $p^{\rm ul}_{|\rm dBm}$ under different scenarios/methods, $\rm NMSE = ||\boldsymbol{h}_{ru}-\hat{\boldsymbol{h}}_{ru}||^2/||\boldsymbol{h}_{ru}||^2$. As we can see from Figure~\ref{fig: NMSE Comparison}, the blue curve, without RI, is the optimum scenario among four different curves, and that is the lower bound for NMSE. The red curve, ignoring RI, is the worst scenario among all, with $p^{\rm ul} = 14~\rm dBm$, the NMSE of the red curve is between $0.04$ to $0.05$. Meanwhile, the purple curve has converged to $\mathrm{NMSE} \approx 0.015$. We also find that FP precoding performs better than null precoding, the yellow curve, which is because although the null precoding has blocked the LoS part of the RI, the NLoS part is still significant compared to local uplink signal power level at the victim BS. FP precoding considers both parts. At $p^{\rm ul} = 14~\rm dBm$, the yellow curve is located around $\mathrm{NMSE} = 0.017$. Besides, we also find that the red curve changes slope twice; this is because we can treat the RI as a two-fold interference source, LoS and NLoS sources, which are independent to each other.
	
	\begin{figure}[t]
		\centering
		\includegraphics[width=1\linewidth]{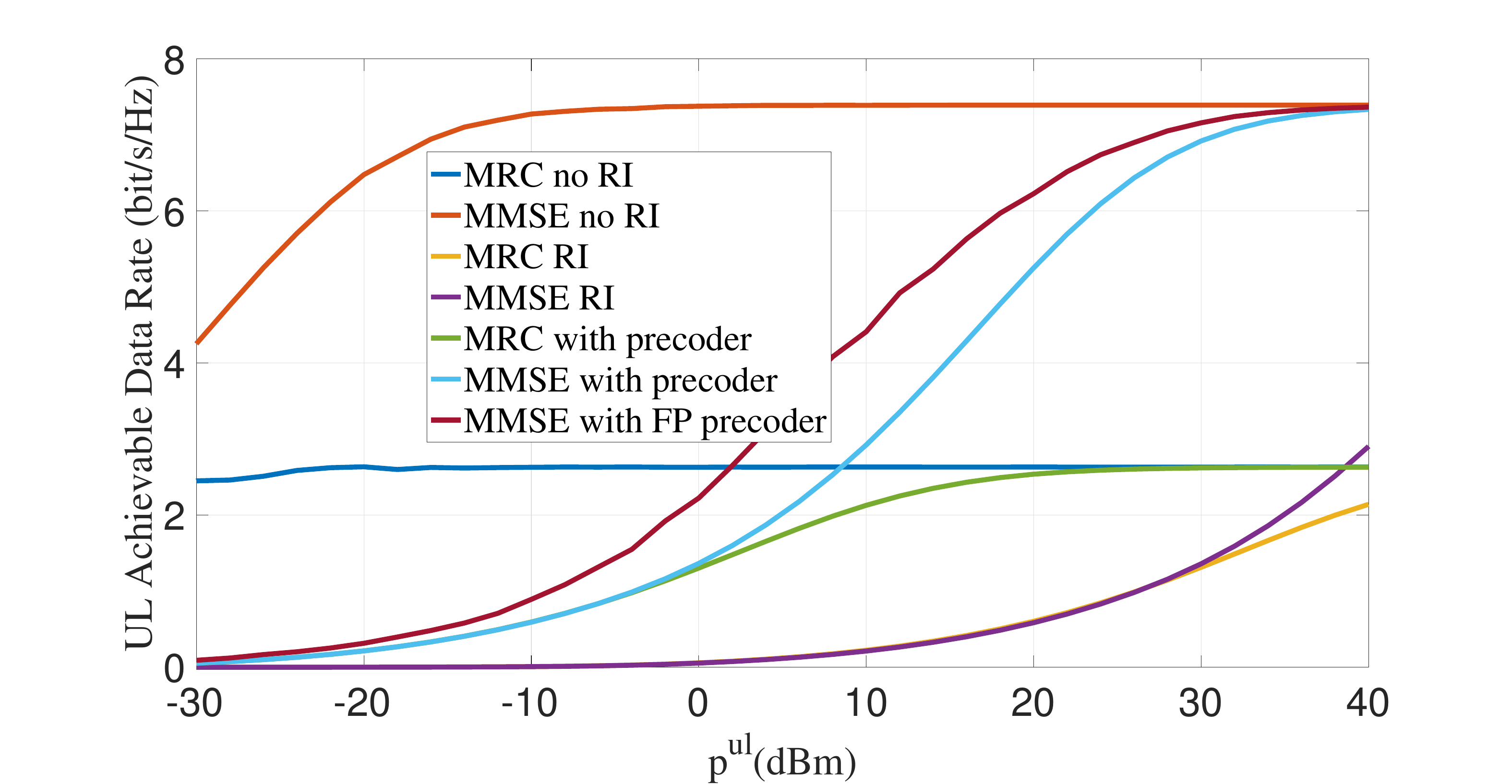}
		\vspace{-1.5em}
		\caption{UL AR vs Transmit Power Comparison}
		\label{fig: UL AR Comparison}
		\vspace{-1em}
	\end{figure}

        Figure \ref{fig: UL AR Comparison} compares uplink achievable rates (AR) for different combiners. Among MRC schemes, no RI yields the highest AR, followed by null precoding, with MRC performing the lowest. The four MMSE-based cases achieve superior AR due to effective interference suppression.
	% Figure \ref{fig: UL AR Comparison} compares uplink achievable rates (AR) under different combiners. The dark blue, green, and yellow curves represent no remote interference (RI), null precoding, and ignoring RI (MRC), respectively. As expected, no RI performs best, ignoring RI performs worst, and null precoding lies in between. The remaining four cases use MMSE combiner, which outperforms the others due to its superior interference suppression capability.
	
	We can see that with no RI, the red curve is the optimum, which can be treated as our upper bound of UL AR. Meanwhile, ignoring RI, the purple curve, which is the worst among these four plots. Once again, the FP precoder performs better than the null precoder, e.g., at $p^{\rm ul} = 14~\rm dBm$, the UL AR using FP precoder is around $5.8$~bit/s/Hz, while using null precoder is around $4.75$~bit/s/Hz, which is a significant improvement. Although the UL AR using FP precoder still has a significant gap compared to the no RI case, it has provided a significant improvement compared to ignoring RI case, which locates around $0$~bit/s/Hz at $p^{\rm ul} = 14~\rm dBm$. Technically, this means that the UEs can transmit at low power regimes that were previously impractical due to the presence of RI. This is expected to enhance the battery life of the victim UEs, contributing to improved energy efficiency and prolonged operational time.

\begin{figure}[t]
		\centering
		\includegraphics[width=1\linewidth]{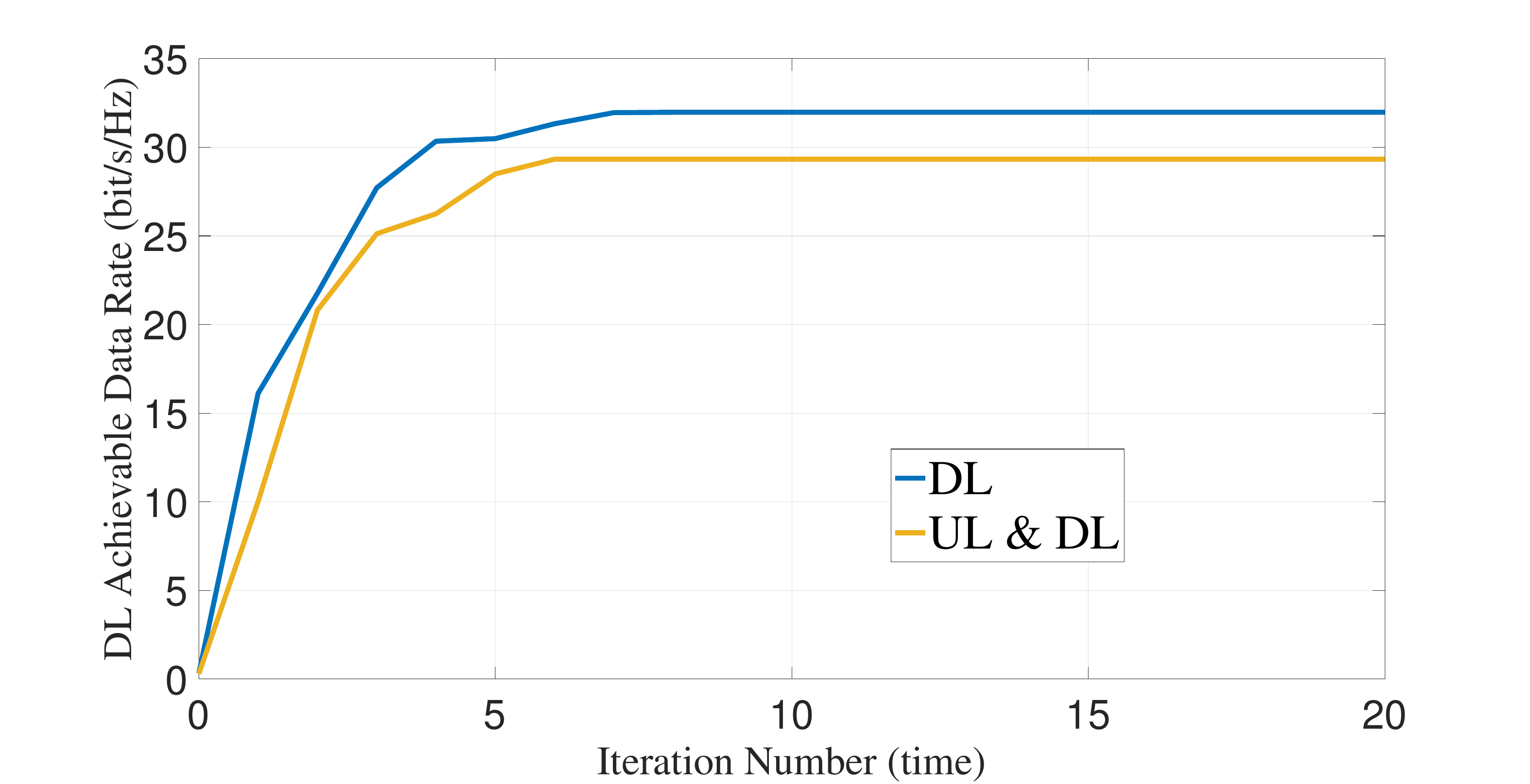}
		\vspace{-1.5em}
		\caption{DL AR vs Iteration Time with FP Precoder}
		\label{fig: DL AR Comparison}
		\vspace{-1.5em}
	\end{figure}

	Finally, we simulate the DL AR using FP vs the iteration time. In Figure~\ref{fig: DL AR Comparison}, we compare the DL AR using different objective functions. The one that considers both the victim UL AR and aggressor DL AR is shown in~\eqref{eq: final} and plotted using the red plot in the figure. Another result that considers the aggressor DL AR only as shown in the second summation term only of~\eqref{eq: final} and plotted using the blue curve in the figure. As expected, if we look at DL AR only, then the blue curve should be higher than the red curve. Since considering the UL AR at the victim side would limit the degree of freedom chosen by FP. Hence, the algorithm needs to balance between focusing on the direction to local UEs and avoiding entering into the troposheric ducting layer. Fortunately, the performance is degraded by only $5$~bit/s/Hz, which is acceptable.
	
	\section{Conclusion}\label{section: conclusion}
	In this paper, we analyzed the impact of RI and proposed two methods to mitigate it, null precoding and FP precoding. The simulation results validate the performance improvement using our proposed methods, specifically, our method results in more accurate CE and higher victim UL AR. We found that although both methods are able to suppress the RI significantly, the FP method always outperforms the nulling method. More importantly, it can keep the DL AR at aggressor side at an acceptable level with acceptable loss. 
	
	Finally, as the distance between the two cellular systems causing RI is large, it is important to limit frequent information exchange between the BSs, e.g., information to notify the aggressor about the combiner and CSI available at victim BS ($\boldsymbol{c}_{ru}$ and $\hat{\boldsymbol{h}}_{ru}$). For future research, we plan to investigate this issue by developing new methods that limit information exchange between the base stations.
	
	%The path loss and shadowing effect between the BS $r$ and UE $u$ at the victim side both vary quite slowly, so the information exchange for these two varying parameters is reasonable. However, the cost for it can be higher than we expect, then finding the average of these two parameter would help this project to be more practical.

\end{document}